\documentclass[aps,
 amsmath,amssymb,
 prl,
 reprint,
 superscriptaddress
 ]{revtex4-2}

\usepackage{graphicx}
\graphicspath{{img/}}
\usepackage{dcolumn}
\usepackage{bm}
\usepackage{hyperref}
\usepackage{cleveref}

\usepackage{multirow}
\usepackage{amsmath,amssymb,amsfonts}
\usepackage{amsthm}
\usepackage{mathrsfs}
\usepackage[title]{appendix}
\usepackage{xcolor}
\usepackage{booktabs}
\usepackage[separate-uncertainty=true]{siunitx}

\usepackage[utf8]{inputenc}
\usepackage[T1]{fontenc}

\begin{document}


\title{Tunable coherence laser interferometry: demonstrating 40dB of straylight suppression and compatibility with resonant optical cavities}


\author{Daniel Voigt}
\email[]{daniel.voigt@uni-hamburg.de}

\author{Leonie Eggers}
\affiliation{Institute for Experimental Physics, 
University of Hamburg, 
Luruper Chaussee 149,
22761 Hamburg,
Germany}

\author{Katharina-Sophie Isleif}
\altaffiliation{Formerly: Max Planck Institute for Gravitational Physics (Albert Einstein Institute), 
Leibniz University Hannover, 
Callinstrasse 38, 
30167 Hannover, 
Germany}
\affiliation{Institute for Automation Technology, 
Helmut Schmidt University, 
Holstenhofweg 85, 
22043 Hamburg, 
Germany}

\author{Sina M. Koehlenbeck}
\altaffiliation{Formerly: Max Planck Institute for Gravitational Physics (Albert Einstein Institute), 
Leibniz University Hannover, 
Callinstrasse 38, 
30167 Hannover, 
Germany}
\affiliation{E.L. Ginzton Laboratory, 
Stanford University, 
348 Via Pueblo, 
Stanford 94305, 
CA, United States}

\author{Melanie Ast}
\affiliation{Max Planck Institute for Gravitational Physics (Albert Einstein Institute),
Leibniz University Hannover, 
Callinstrasse 38, 
30167 Hannover, 
Germany}

\author{Oliver Gerberding}
\email[]{oliver.gerberding@uni-hamburg.de}
\affiliation{Institute for Experimental Physics, 
University of Hamburg, 
Luruper Chaussee 149,
22761 Hamburg,
Germany}

\date{\today}

\begin{abstract}
A major limitation of laser interferometers using continuous wave lasers are parasitic light fields, such as ghost beams, scattered or stray light, which can cause non-linear noise. This is especially relevant for laser interferometric ground-based gravitational wave detectors. 
Increasing their sensitivity, particularly at frequencies below 10\,Hz, is threatened by the influence of parasitic photons. These can up-convert low-frequency disturbances into phase and amplitude noise inside the relevant measurement band.
By artificially tuning the coherence of the lasers, using pseudo-random-noise (PRN) phase modulations, this influence of parasitic fields can be suppressed. As it relies on these fields traveling different paths, it does not sacrifice the coherence for the intentional interference.
We demonstrate the feasibility of this technique experimentally, achieving noise suppression levels of 40\,dB in a Michelson interferometer with an artificial coherence length below 30\,cm. We probe how the suppression depends on the delay mismatch and length of the PRN sequence.
We also prove that optical resonators can be operated in the presence of PRN modulation by measuring the behavior of a linear cavity with and without such a modulation. By matching the resonators round-trip length and the PRN sequence repetition length, the classic response is recovered.
\end{abstract}


\maketitle

\section{Introduction}\label{sec:introduction}
Laser interferometers are widely used metrological tools in fundamental research, various measurement instruments and industry applications. Interferometers that use continuous wave lasers with high stability profit from the associated strong spatial coherence to generate interference patterns in various scenarios. This includes interferometers with very long arms or high Finesse optical resonators where thousands of round-trip fields are interfered to achieve, for example, power enhancement. A prime example using both these techniques are laser interferometric, ground-based gravitational wave detectors. These kilometer-long interferometers measure relative arm-length strain with an integrated sensitivity around $10^{-20}$ at frequencies between 10\,Hz and 10\,kHz to detect fluctuations of space-time itself. 
The current network of observatories \cite{LVK_Network} consisting of the Advanced LIGO, Advanced VIRGO \cite{VIRGO}, KAGRA \cite{KAGRA} and GEO600 \cite{GEO600} detectors steadily increases the rate of detected events through ongoing improvements and updates of the interferometers. Even further increases in sensitivity will be realized in the next generation of observatories like the Einstein Telescope (ET) \cite{ET} and Cosmic Explorer (CE) \cite{CosmicExplorer,CosmicExplorer2}, where the former specifically aims to extend the sensitivity down to 3\,Hz. 

However, these large interferometers, along with smaller ones used, for example, for rotation measurements \cite{ringlaser_Korth2015,ringlaser_igel2021, ringlaser_schreiber2023}, not only profit from the large laser coherence. Any photons sent into the interferometers that are not following the nominal beam trajectory but still couple back into the sensitive path can lead to measurement disturbances. These parasitic interferences are a well-known and often critical challenge in precision laser interferometry. 

In ground-based gravitational wave detectors, reducing the influence of these unwanted light fields, often called ghost beams, straylight or scattered light, belongs to the larger efforts of mitigating noise sources that limit the sensitivity of the observatories \cite{martynov2016}.
Currently, there are several techniques employed to mitigate the impact of scattered light, many of which are also standard in other precision interferometers, including baffles~\cite{Soni2024}, beam dumps and reducing scattering sources and their relative motion driven by e.g. seismic or acoustic motion \cite{soni2021,nguyen2021}. In other studies the scattered light is phase modulated after leaving the main beam, changing its dynamics directly \cite{schilling1981,schnupp1985,luck2008} to reduce the coupling into the measurement band. Subtraction of scattered light effects has also been studied in different schemes, using existing auxiliary photodetector signals \cite{Was2021, Acernese2022} or the readout of additional light quadratures at the nominal output \cite{ast2016}.

However, for the sensitivity increase aimed for in future observatories, even single photons scattered back into the readout will be troublesome \cite{chua2014,ottaway2012,andres-carcasona2023}. Thus, the suppression of parasitic fields is a permanent and major effort in precision interferometers and has a major impact on their complexity in both design and implementation. 

This motivates our study of a new, fundamental solution to the suppression of parasitic light fields in laser interferometers, specifically for future ground-based gravitational wave detectors and, generally, for all interferometers limited by such effects.

The concept we call \textit{tunable coherence} aims to carefully control and break the coherence of laser light. The coherence and therefore interference are reduced to a minimal value for all unintended optical paths while maintaining it for the intended path. This effectively creates a pseudo-white-light interferometer which is inherently insensitive to parasitic light fields having a relative delay to the intended path.
We achieve this with \si{\giga\hertz}-phase-modulation following a pseudo-random-noise (PRN) binary sequence applied to the laser light \cite{voigt2023}. The sequence, consisting of deterministically but seemingly randomly generated chips of logical zeros and ones, flips the light phase by \SI{180}{\degree}. This concept was already briefly considered in the context of gravitational wave interferometers \cite{dewey1986}, but deemed unfeasible, mostly due to technical limitations in available components. However, advances in e.g. electro-optical components and successes with multiplexed readout in digitally enhanced interferometry \cite{sutton2012,shaddock2007} have changed that. It now seems feasible and relevant to investigate the application of tunable coherence to the main science laser of gravitational wave detectors and other precise laser interferometers. 

\begin{figure}[!t]
    \includegraphics[width=0.49\textwidth]{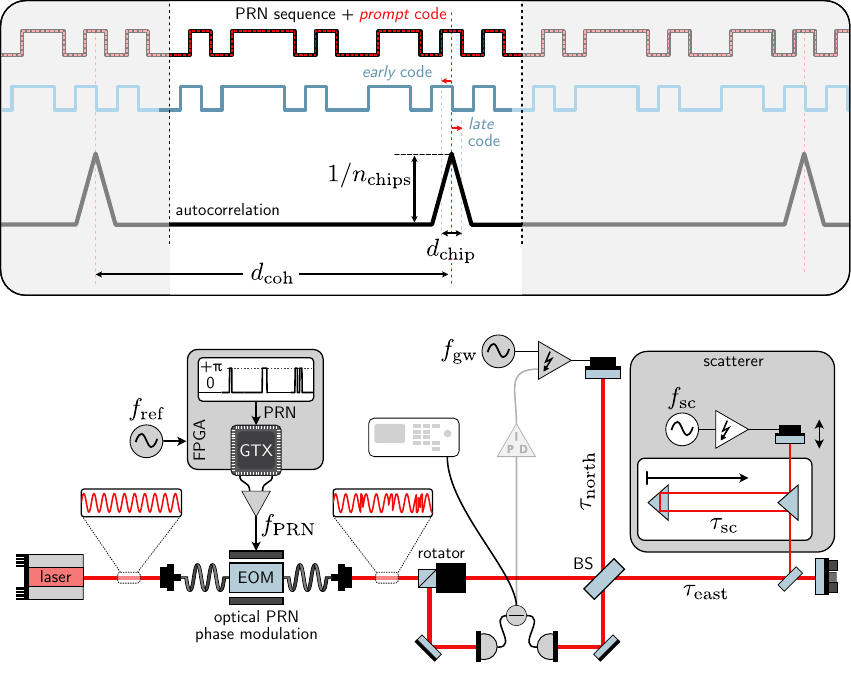} 
    \caption[]{The properties of tunable coherence and the PRN modulation are sketched in the top figure. The sketch below depicts the experimental setup with the fiber-coupled electro-optic modulator (EOM) modulating the PRN-sequence onto the laser at $f_{\text{PRN}}$ and the Michelson interferometer locked at mid-fringe. The north-arm was used to actuate the phase between the two arms and locking the interferometer, in the east-arm some light was reflected out of the arm and coupled back after being reflected from another actuated mirror to simulate scattered light coupling. This scattered light picked up an additional delay $\tau_{\text{sc}}$ relative to the equal arm-delays $\tau_{\text{north}}$ and $\tau_{\text{east}}$.}  
    \label{fig:experiment}
\end{figure}

With using tunable coherence, two new length scales are introduced which are also sketched in the top part of Figure~\ref{fig:experiment}:

The remaining coherence length set by the length of the chips, $d_{\text{chip}}$, that form the binary modulation sequence. This optical length given by the inverse of the PRN-modulations frequency, $f_{\text{PRN}}$, and the speed of light, $c$, is a measure for the minimum needed relative delay to suppress interference.
The width of remaining coherence is thus double $d_{\text{chip}}$.

And the length until the coherence is restored after delaying one beam by a whole PRN-sequence. 
This so-called \textit{recoherence length}, $d_{\text{coh}}$, is determined by the integer length of the sequence, typically called the number of chips, $n_{\text{chips}}$, and their length.
Coherence is restored for all integer multiples of this length.

Both lengths can be written as
\begin{equation}
    \label{eq:lengths}
    d_{\text{chip}} = \frac{c}{f_{\text{PRN}}} \qquad \text{and} \qquad d_{\text{coh}} = n_{\text{chips}}\cdot d_{\text{chip}}.
\end{equation}

In our scheme the demodulation is done optically by interfering the modulated light beams with a path length difference close to zero but significantly smaller than $d_{\text{chip}}$.
The use of photodetectors with bandwidths below the sequences repetition rate ($f_{\text{PRN}} /n_{\text{chips}} $) then removes the modulation signal. Thus, the photodetector signals are effectively the same as in non-phase-modulated interferometers, assuming all nominal interferences are delay matched.

Since the main application we focus on are laser interferometers based on an extended Michelson topology using additional optical resonators, we present here experiments to demonstrate the ability of tunable coherence to suppress the influence of parasitic beams in a classic Michelson interferometer and demonstrate the compatibility of tunable coherence with an optical resonator.

\section{Results}\label{sec:results}
For the PRN-modulation in both experimental setups several PRN-sequences were stored on a field-programmable-gate-array (FPGA) and modulated onto the laser using a fiber-coupled waveguide electro-optical modulator (EOM) with a maximal bandwidth of \SI{20}{\giga\hertz}. The EOM driving voltage was manually adjusted by minimizing the scattered light tone in a live-spectrum. Details about the modulation setup can be found in Methods.

\subsection{Tunable coherence in a Michelson interferometer}\label{sec:michelson}
The first experiment consisted of a Michelson interferometer with matched equal arm lengths around \SI{1}{\meter}, as shown in Figure~\ref{fig:experiment}. To create a parasitic beam path, we placed a low power reflectivity ($R\approx 0.2$) mirror in one arm to couple some light out and back in. This beam path could be actuated with a piezo-actuated mirror and its optical length changed with a delay-line. Details of the setup are described in Methods.

We injected two sine-signals, one simulating a gravitational wave at $f_{\text{gw}} = \SI{172.4}{\kilo\hertz}$ with a piezo-actuator in the north-arm directly into the interferometer and one at $f_{\text{sc}} = \SI{170}{\kilo\hertz}$ to modulate the phase of the parasitic beam simulating scattered light 
coupled into the east-arm.

As the coupling of the parasitic beam into the interferometer readout is non-linear \cite{schilling1981} the strength of its introduced phase error depends on the DC-phase relative to the main light field, which in turn fluctuates slowly. To ensure we always observed the maximum noise coupling we therefore ramped the piezo-actuator slowly through several fringes over each measurement duration. 
This guaranteed that each measured time series always contained the strongest possible coupling, thus allowing for comparison between data taken asynchronously.

Data was recorded in time series either with or without the PRN-modulation. The sampling rate and duration were adjusted to ensure each sample contained at least one full PRN-sequence. This is crucial to avoid artifacts of the modulation showing in the data and it is advisable to have even multiple full sequences in an average to reduce any spurious artifacts~\cite{sibley2020}.

\begin{figure}[!t]
    \centering
    \includegraphics[width=0.48\textwidth]{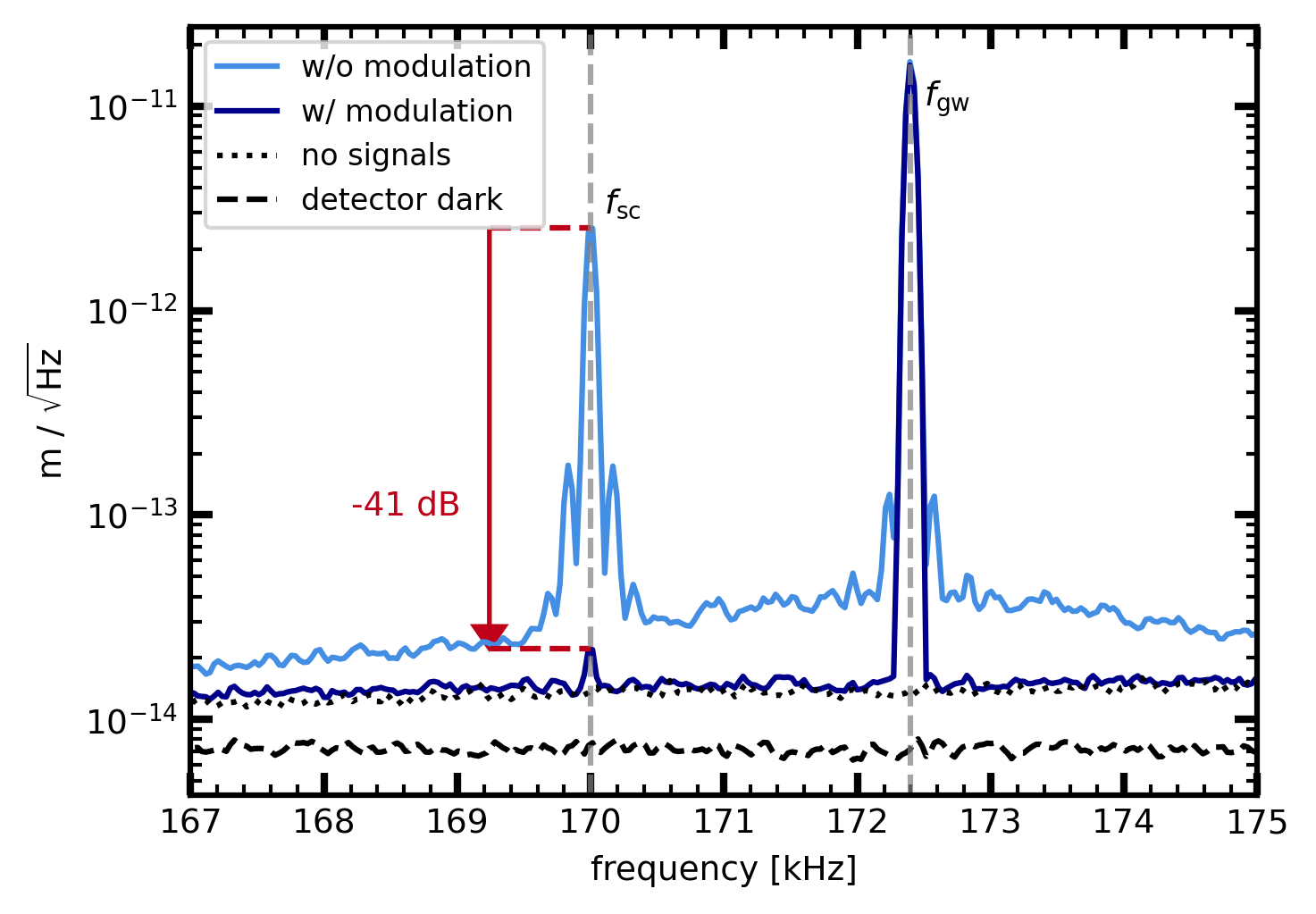}
    \caption{Amplitude spectral density of the interferometer output without and with PRN-modulation using a 255-chips long sequence. The peak at \SI{170}{\kilo\hertz} corresponds to the injected scattered light modulation, while the peak at \SI{172.4}{\kilo\hertz} is from a signal injected directly into the north arm of the interferometer, simulating a gravitational wave. Using the PRN-modulation, the scattered light peak was reduced by \SI{41}{\decibel}.}
    \label{fig:suppression8bits}
\end{figure}

From this data, several spectra were computed. Details of the recording and treatment of data can be found in Methods. An example measurement and its spectra are shown in Figure~\ref{fig:suppression8bits} for a PRN-modulation with a length of 255 chips. This measurement demonstrated a noise suppression of about \SI{41}{\decibel} for the parasitic tone while maintaining the desired simulated gravitational wave signal. Moreover, also the surrounding noise floor and side-lobes of this signal were reduced, which we identified to be caused by the parasitic beam path in the interferometer by blocking this path. The measurement also shows that with PRN-modulation present the original noise floor could be recovered, except for a residual coupling of the parasitic tone. 

\begin{figure*}[!t]
    \centering
    \includegraphics[width=0.95\linewidth]{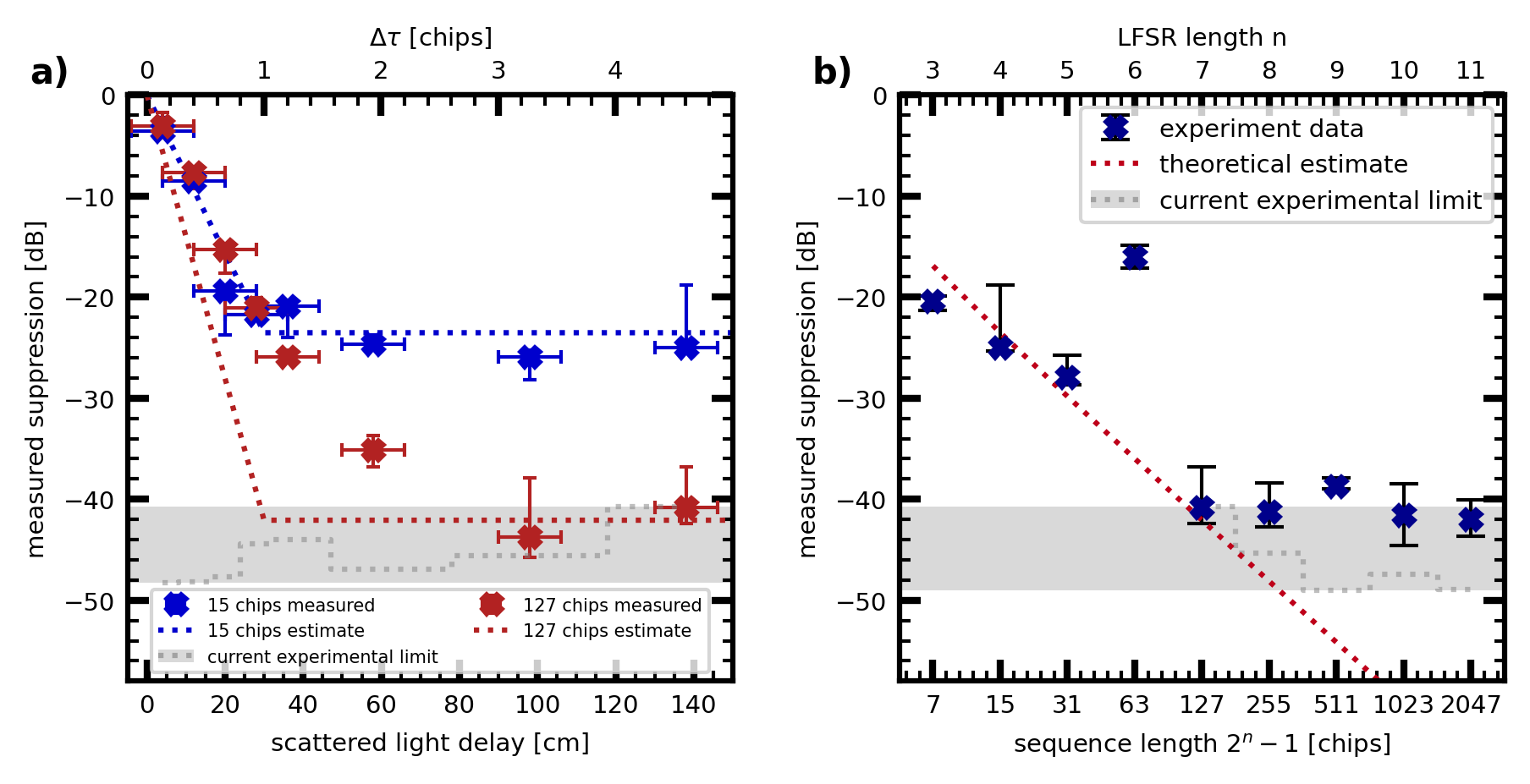}
    \caption{Measured suppression of the scattered light depending on the delay of the scattered light relative to the interferometer arms (figure a)
    and on the length of the used PRN-sequence (figure b).\\
    (a) The suppression for the 15-chip sequence follows the theoretical estimates, while the 127 chip sequence is not dropping of as sharply as
    theoretically predicted. The latter still achieves the maximum expected suppression for longer spatial delays.\\
    (b) The dependence of the suppression on the PRN-sequence length mostly follows the theoretical estimate until the maximum suppression level achieved in our experimental set-up is reached. Only the 63-chip sequence shows an unexpectedly low suppression.
    }
    \label{fig:suppressionVsParameter}
\end{figure*}

We investigated the dependency of this suppression on two parameters, the relative delay of the parasitic beam, tuned with the delay-line, and the length of the PRN-sequence, $n_{\text{chips}}$. The results are shown in Figure~\ref{fig:suppressionVsParameter}.

The suppression for different scatter delays (Figure~\ref{fig:suppressionVsParameter}a) is shown for two sequence lengths, 15 and 127 chips. The data for the 15~chips sequence follows the theoretical limitations given by the minimal coherence and its dependency on the chip length. It reaches its maximum possible suppression after a delay exceeding $\SI{30}{\centi\meter}$ or $\SI{1}{\nano\second}$, the length of one chip.

For the 127~chips long sequence, a fade out of the curve compared to the sharp theoretical estimate can be observed. The maximum suppression is still achieved, but only for longer delays than calculated. The estimate is based on an assumed perfectly binary phase modulation. However, any real modulation shows effects of limited bandwidth and non-flat transfer-functions, which will deform the ideal shape and broaden the coherence length.

The results on measuring the maximum suppression depending on PRN-sequence length are shown in Figure~\ref{fig:suppressionVsParameter}b. 
The dependence roughly follows the theoretical estimate $20\log_{10}(\frac{1}{l_{seq}})$ given by the auto-correlation function of the sequence \cite{voigt2023} with the 63 chips sequence being a strong outlier. It then saturates at our current experimental limit of suppression levels around \SI{40}{\decibel}. This level closely coincides with the limit given by the strength of the parasitic tone in comparison to the measurement noise floor, which fluctuates between measurements (indicated as dashed gray lines in Figure~\ref{fig:suppressionVsParameter}). However, for some measurements with long sequences we still not fully suppress the parasitic tone, which indicates another limitation lying in the same range. Nevertheless, except for the 63~chips sequence, the data follows expectations and we reach a maximum of \SI{41.7}{\decibel} for the 2047~chips long sequence. 

\subsection{Tunable coherence in a cavity}\label{sec:cavity}
\begin{figure}[!t]
    \centering
    \includegraphics[width=0.49\textwidth]{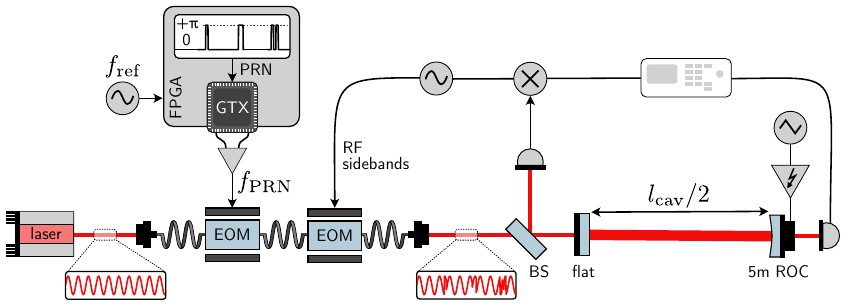}
    \caption[]{ The sketch depicts the second experimental setup with the fiber-coupled EOM modulating the PRN-sequence onto the laser at $f_{\text{PRN}}$, another EOM for RF sideband generation and the light coupled into a linear optical cavity. The actual cavity was folded, but is depicted here on one axis for simplicity. The round-trip length of the cavity $l_{\text{cav}}$ was fixed in each measurement, while the PRN-sequence repetition length was tuned by adjusting the reference clock frequency $f_{\text{ref}}$ of the FPGA that is proportional to the PRN-frequency $f_{\text{PRN}}$.}
    \label{fig:experiment2}
\end{figure}

The second experiment consisted of a folded, linear cavity in which we injected the PRN-modulated light, as sketched in Figure~\ref{fig:experiment2}. The cavity was tunable in microscopic length and read out in transmission and reflection. A second EOM was introduced for additional RF phase modulation.

The cavities round-trip length, $l_{\text{cav}}$, was chosen to match the recoherence length, $d_{\text{coh}}$ of a 15~chips long sequence at \SI{1}{\giga\hertz} PRN-frequency. In this setup, we matched the recoherence length to $l_{\text{cav}}$ by tuning the PRN-frequency. However, one could also fine-tune the macroscopic cavity length to a given $d_{\text{coh}}$, preferably with a fine-adjustable opto-mechanic to sustain optical contrast. By tuning the PRN-frequency, assuming it is well known, one can also precisely measure the absolute cavity length, in multiples of $d_{\text{coh}}$. Here, the length deviated \SI{3.388}{\milli\meter} or \SI{0.15}{\percent} from the geometrical measurement seen in the necessary detuning from \SI{1}{\giga\hertz} PRN-frequency. 

Additionally to this case, data was also taken for the cavity length matching half the recoherence length ($l_{\text{cav}} / d_{\text{coh}} = 0.5$) and twice the recoherence length ($l_{\text{cav}} / d_{\text{coh}} = 2$). For the former, we expect half the power build-up inside the cavity, since only every second round-trip will create a constructive interference~\cite{voigt2023}. For the latter, it was necessary to physically shorten the cavity due to limitations in the tuning range of the PRN-frequency. Detailed descriptions of the setup can be found in Methods.

\begin{figure}[!t]
    \centering
    \includegraphics[width=0.48\textwidth]{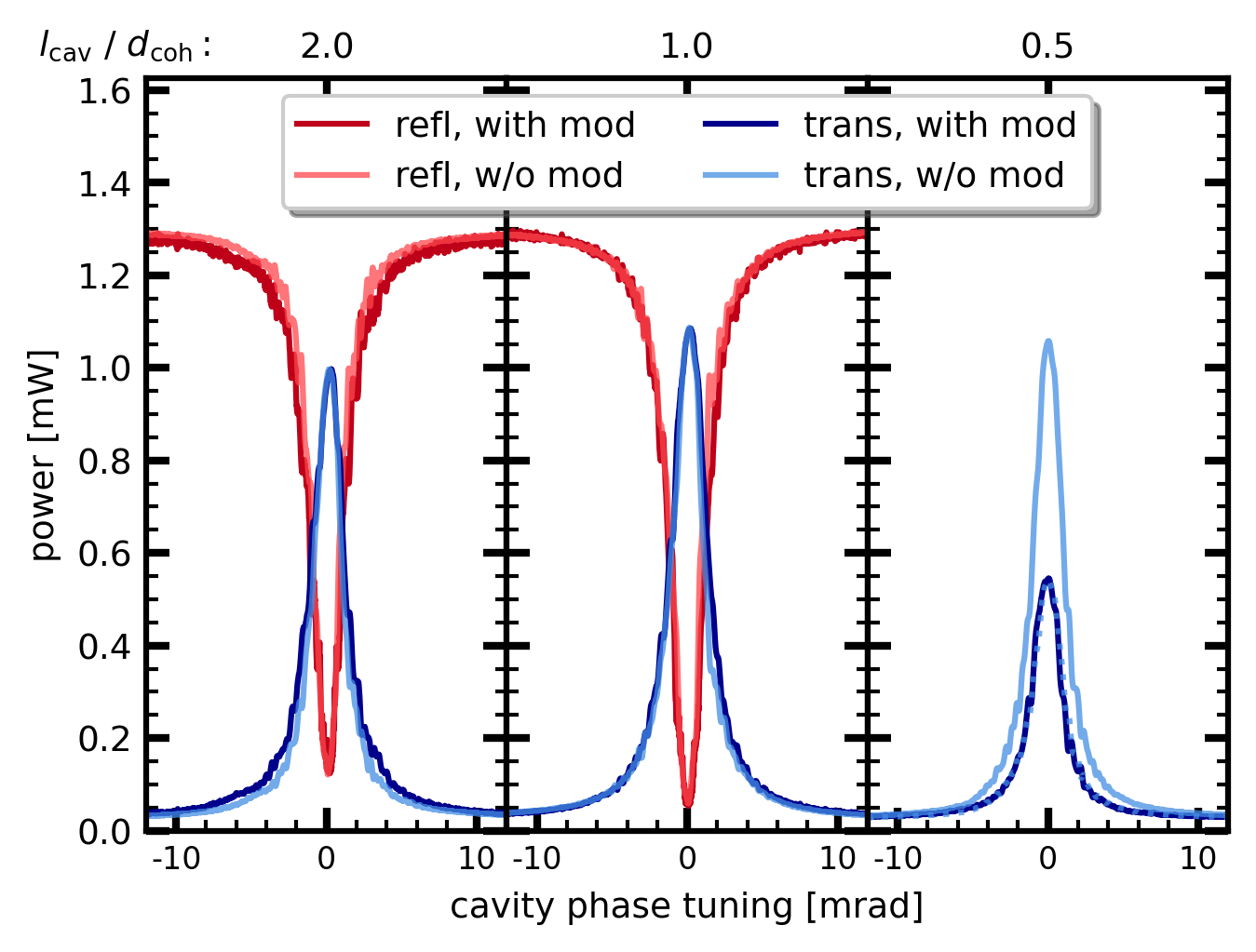}
    \caption{Measured cavity transmission and reflection without and with modulation using three different sequence lengths (left: 7 chips, middle: 15 chips, right: 31 chips). The longer \SI{2.252}{\meter} cavity allowed for the 15-chips long sequence to be resonant with a real modulation frequency of $f_\mathrm{PRN}\,=\,\SI{0.998495}{\giga\hertz}$ and the 31-chips long sequence at $f_\mathrm{PRN}\,=\,\SI{1.03177}{\giga\hertz}$. The 7-chips long sequence was resonant in the shorter \SI{2.100}{\meter} long cavity at $f_{PRN}\,=\,\SI{0.99894}{\giga\hertz}$. As this was a physically different cavity, the finesse is slightly worse due to alignment inaccuracies.}
    \label{fig:cavityscan}
\end{figure}

We first measured the transmission and reflection during a scan over the resonance for each case with and without PRN-modulation. The measurements, with example time-series shown in Figure~\ref{fig:cavityscan}, indicate no measurable differences in the resonant behavior for integer relations between the lengths. Only for the half-integer relation we find the expected reduction in transmitted power~\cite{voigt2023}.

\begin{figure*}[!t]
    \centering
    \includegraphics[width=0.95\textwidth]{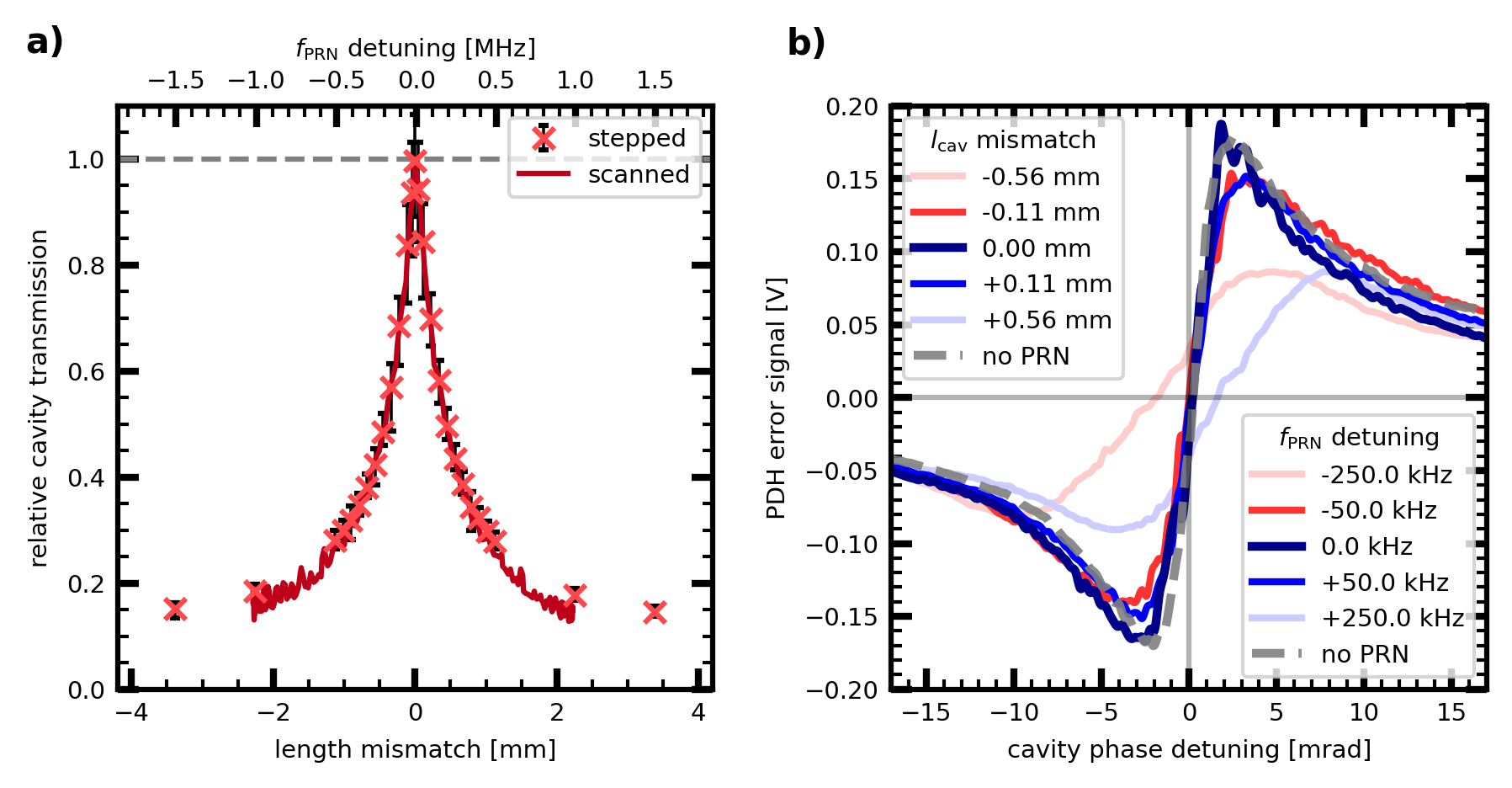}
    \caption{Cavity behavior measured with different detunings of the PRN-frequency $f_{\text{PRN}}$ and thus mismatch between round-trip and recoherence length. Figure a shows the relative transmitted power as a measure for power build-up inside the cavity and figure b shows the resulting Pound-Drever-Hall error signals measured in reflection.\\
    (a) Step-wise detuning and scanning the PRN-frequency while the cavity is locked on microscopic resonance show a new resonance depending on macroscopic length matching to the PRN-sequence having a FWHM of around \SI{710}{\micro\meter} for \SI{1}{\giga\hertz} PRN-modulation and a cavity Finesse of around 696.\\
    (b) PDH error signals measured with a matched and only slightly ($\pm$\SI{50}{\kilo\hertz}) detuned PRN-sequence showing no meaningful deviation from the case without tunable coherence while the further ($\pm$\SI{250}{\kilo\hertz}) PRN-modulation deteriorates the error signal significantly.}
    \label{fig:cavityresonance+PDH}
\end{figure*}

Next, we repeated this measurement for different detunings of the PRN-frequency, equaling length mismatches between cavity and PRN-sequence, and measured the transmitted power relative to the case without PRN-modulation.
Additionally to this stepped measurement, we locked the cavity to resonance and scanned the PRN-frequency between $\pm$\SI{1}{\mega\hertz} detuning, continuously recording the transmitted power. Both measurements are displayed in Figure~\ref{fig:cavityresonance+PDH}a for the case where $l_{\text{cav}}$ and $d_{\text{coh}}$ were equal. Both other cases also showed the expected behavior.

From this measurement it is possible to determine a new full-width-half-maximum (FWHM) for the cavity being resonant depending on its macroscopic length additionally to the microscopic phase tuning. In our case with \SI{1}{\giga\hertz} modulation in the given cavity, this FWHM lies around \SI{710}{\micro\meter}. It becomes clear that for parasitic beams traveling with round-trips not matching the cavity mode, 
especially bouncing back from some exterior surface, this reduces the resonance. From previous simulations~\cite{voigt2023} and this experiment, we determine that this FWHM introduced by \textit{tunable coherence} depends on PRN-frequency $f_{\text{PRN}}$ as it determines the remaining coherence length (see eq.~\ref{eq:lengths}) -- and the cavities Finesse due to the number of round-trips summing up a mismatch.

Lastly, the effect of tunable coherence on RF sidebands was measured by adding a phase modulation at $f_\mathrm{PDH}\,=\,\SI{5}{\mega\hertz}$. The signal was demodulated in reflection to generate the well known Pound-Drever-Hall (PDH) error signal. We again scanned the cavity over its resonance and characterized the PDH error signal for different detunings of the PRN-modulation, as shown in Figure~\ref{fig:cavityresonance+PDH}b. 
We see that for perfect matching between sequence and cavity the error signal is preserved. Futhermore, for slight detunings of the lengths neither the zero-crossing, nor the slope of the error signals change in a meaningful way, indicating general compatibility of PDH and tunable coherence. Only for larger mismatches we find a significant deviation making the error signal unusable. Especially for detunings reaching minimal transmitted power, no usable error signal could be obtained. However, an acquired lock withstands farther detuning as seen in Figure~\ref{fig:cavityresonance+PDH}a where we detuned the PRN-frequency in a locked cavity by up to \SI{1}{\mega\hertz}.

\section{Discussion}\label{sec:discussion}
Our measurements prove that tunable coherence is a viable and ready technique to reduce the impact of parasitic beams and scattered light in laser interferometers. The spatial coherence can be reduced to the 10\,cm scale and parasitic phase noise suppression was verified up to \SI{41}{\decibel} without introducing any new noise within our current measurement precision. Optical resonators and PDH-locking were also demonstrated to be compatible with the technique. These first measurements open a promising avenue to investigate the technique further and push for even shorter coherence lengths and higher noise suppression.

Compared to other experiments, that achieved around one order of magnitude suppression without further post-processing~\cite{luck2008}, the demonstrated suppression here considerably shifts the state-of-the-art while also removing the need to identify individual scatter-sources. 
As phase errors $\varphi_{\text{err}}$ introduced by straylight scale with the spurious power $P_{s}$ as
\begin{equation}
    \label{eq:effect}
    \varphi_{\text{err}} \propto \sqrt{P_{s}},
\end{equation}
a suppression of \SI{40}{\decibel} means an increase in tolerable parasitic beam power by a factor 10,000. This would significantly simplify efforts to reduce straylight in, for example, ground-based gravitational wave detectors. 

The current limitation to the suppression measurement is close to our experimental limitations to generate parasitic tones, however, we suspect other effects to also be relevant in this range, mainly deviations from perfect phase-flips in the PRN-modulation. Only for ideal \SI{180}{\degree} phase-flips maximal suppression of the carrier tone can be assumed. Otherwise it can still contain significant power, reducing the achieved suppression as it competes with the PRN-modulated fields. 
While we optimized the modulation depth in the experiment, it was not permanently measured or controlled and expected to drift.
With the ability to handle these effects, one could push for stronger suppression using longer PRN-sequences in the future. An option is to actively stabilize the PRN-modulation depth, which is possible but requires a complex control scheme~\cite{Yang2011}.
Nevertheless, as the suppression of effects from spurious beams scales linearly with the number of chips in the sequence $n_{\text{chips}}$~\cite{shaddock2007}:
\begin{equation}
    \label{eq:reserr}
    \varphi_{\text{res,err}} = \frac{ \varphi_{\text{err}} }{n_{\text{chips}}}, 
\end{equation}
with $\varphi_{\text{res,err}}$ as residual phase error, the tolerable parasitic beam power scales quadratic with sequence length, strongly motivating longer sequences.

The deficiency in reaching full suppression after one chip-length relative delay also indicates an influence of limited bandwidth in the PRN generation and application. 
This imperfection degrades the auto-correlation-function of the used sequence, limiting the reachable suppression for delays close to the length of one chip. Optimizing electronics and cabling will help to push this limit further, with various high-speed equipment available from telecommunication technology. 
Modern phase modulators are reaching several GHz of bandwidth, allowing the coherence lengths to be reduced to the centimeter scale. Moreover, such short chips enable one to use sequences that significantly suppress unwanted fields while their recoherence lengths are still short enough to realistically fit optical resonators as demonstrated here. Inside those resonators, the effects of hundreds of round-trips needing to interfere further reduces the coherence length, in our demonstration below \SI{1}{\milli\meter}. Additionally, using  PRN-modulation in the \si{\giga\hertz}-range ensures good separation of RF signal modulations from the sequence repetition rate for moderate sequence lengths. 

However, we see some open challenges in our data that need addressing. One sequences does not perform as expected while the very short 7~chips long sequence often outperforms expectations. This is currently not understood, but we 
verified the correctness of the used m-sequences in various ways, leaving it open to be another, yet unknown effect. 
Additionally, we have realized the current proof-of-concept with measurements in the \SI{100}{\kilo \hertz}-range and have yet to show shot-noise limited performance at frequencies down to a few \si{\hertz}, where new effects might arise.

The successful demonstration of tunable coherence in optical resonators indicates that more complex interferometer topologies using several resonators can be operated with this technique. This is especially relevant for the application in gravitational wave detectors, which use not only 4 main resonators in the dual-recycled Fabry-Perot Michelson 
interferometer configuration \cite{DRFPMI_Staley2014} to optimize their quantum noise sensitivity, but also a number of cavities as e.g. mode-cleaners to reduce other technical laser noise. 

Furthermore, the ability of tunable coherence to reduce elastic effects of parasitic light fields in these detectors, causing phase noise via radiation pressure~\cite{canuel2013}, should be demonstrated. While a simple argument can be made that these effects will become a broadband background in the presence of PRN-modulation, experimental verification will be necessary.
Similarly, the compatibility with quantum noise reduction schemes, prominently via squeezed light injection, has to be demonstrated. 

Generally, tunable coherence interferometry has a major implication for the optical layout, requiring all relevant optical lengths to be matched to an integer multiple of the recoherence length. In gravitational wave detectors this is certainly feasible, but complicated by features such as the Schnupp-Asymmetry, an intentional arm-length difference currently used for locking signal-recycling-cavities~\cite{schnupp1988}. 

Ghost beams, scattered light and straylight remain one of the major technical challenges in precision laser interferometers. With tunable coherence we have demonstrated a new way of dealing with this permanent and inconvenient companion that can help pushing for new levels of sensitivity and simpler optical layouts.

\section{Acknowledgements}
\begin{acknowledgments}
    This research was funded by the Deutsche Forschungsgemeinschaft (DFG, German Research Foundation) under Germany's Excellence Strategy---EXC 2121 ``Quantum Universe''---390833306. The authors thank André Lohde for his assistance in setting up the lab infrastructure during his time in the group.
\end{acknowledgments}

\section{End Matter}
\appendix
\section{Methods}\label{sec:methods}
\subsection{PRN modulation generation}
For our table-top experiments we used a non-planar ring oscillator (NPRO) \SI{1064}{\nano\meter} laser (Coherent: Mephisto 500) that was send through a fiber coupled waveguide electro-optical modulator (EOM) (iXblue Photonics: NIR-MPX-LN-20) with a maximal bandwidth of \SI{20}{\giga\hertz}. The PRN sequences were so-called maximum-length-sequences (\textit{m}-sequences) generated by linear-feedback-shift-registers (LFSR) with varying lengths~\cite{alfke1996}, such that we could tune $n_{\text{chips}}$ between 7 and 2047. To reach the desired generation speeds the sequences were stored on a field-programmable-gate-array (FPGA) (AMD Artix-7 in Zynq 7000 XC7Z045 SoC on ZC706 Evaluation Board) and transmitted as differential signal to a dedicated EOM driver (iXblue Photonics (exail): DR-DG-20-HO) using an onboard serial \si{\giga\hertz}-GTX transceiver of the ZC706 evaluation board. The sequences were modulated with a frequency of ${f_{\text{PRN}}} = \SI{1}{\giga\hertz}$.

Interfering the PRN modulated with unmodulated light allowed for in-situ observation of the PRN sequence with a high-speed photo-diode (Thorlabs: RXM42AF, 42 GHz Photoreceiver) while the EOM driving voltage could be optimized by observing the strength of the injected scattered light note in a live-spectrum.

\subsection{Michelson interferometer setup}
The experimental setup to demonstrate straylight suppression with tunable coherence consisted of a Michelson interferometer with equal arm-length of around \SI{1}{\meter}. 
The north-arm contained a piezo-actuated mirror for adjusting the phase, locking the interferometer and simulating (gravitational wave) signals. In the east arm light was split of the interferometer by a low power reflectivity ($R\approx 0.2$) mirror to create a nominal and a parasitic beam path. 
The parasitic beam was phase modulated by another piezo-driven mirror and coupled back via the same low-reflectivity mirror. The delay $\tau_{sc}$ relative to the light in the interferometer could be adjusted with an optical delay-line.

We locked and read out the interferometer at mid-fringe by taking the difference between the light power measured at the anti-symmetric and symmetric port. 
The control-loop for this lock had a unity gain frequency of around \SI{5.5}{\kilo\hertz}, resulting in the interferometer being effectively free running at frequencies above \SI{100}{\kilo\hertz}. At these frequencies we were only limited by residual laser amplitude noise.  
We injected two different sine-signals, one simulating a gravitational wave at $f_{\text{gw}} = \SI{172.4}{\kilo\hertz}$ with the piezo-actuator in the north-arm directly into the interferometer and one at $f_{\text{sc}} = \SI{170}{\kilo\hertz}$ to modulate the phase of the parasitic beam that simulates scattered light 
coupled into the east-arm. As the DC-phase offset between interferometer and scattered light path slightly fluctuated, we additionally ramped it through several fringes at a low frequency during each measurement.

\subsubsection{Recording and treatment of data}
Data was recorded by taking time series either with or without the PRN modulation active. 
These had around four million data points sampled over at least two seconds with a sampling rate of \SI{2}{\mega\hertz}. As the corresponding sampling rate had to be reduced for longer PRN-sequences (511 chips and more) to ensure averaging over at least one full sequence, the sampling duration was increased accordingly to four and eight seconds at \SI{977}{\kilo\hertz} and \SI{488}{\kilo\hertz} sampling frequency respectively.

From the recorded data, several spectra were computed using Welch's method with a Blackman-Haris window and \SI{50}{\percent} overlap. The reached suppression was calculated by comparing the peaks at $f_{\text{sc}}=\SI{170}{\kilo\hertz}$ between spectra calculated over the full time series measured with and without PRN-modulation. The upper and lower limits were calculated by comparing minimum and maximum values of the same peak in the spectra, this time calculated for 30 parts of the same length distributed evenly over the whole recorded time series.

\subsection{Linear cavity setup}
The experimental setup to demonstrate compatibility of tunable coherence and optical resonators consisted of a folded, linear cavity. This cavity was microscopically tuned in length with a piezo-actuated mirror, the folding mirror. It was setup using two mirrors with \SI{99.5}{\percent} power reflectivity as input and end mirrors, the input mirror was flat, the end mirror had a radius of curvature of \SI{5}{\meter}. Two photodetectors in transmission and reflection measured the transmitted and reflected laser power, respectively. The Finesse of the cavity was around 696.

For an additional RF phase modulation of the injected light field a second fiber-coupled EOM (iXblue Photonics: NIR-MPX-LN-0.1) with a bandwidth of \SI{150}{\mega\hertz} was introduced. This EOM was driven by the a second dedicated EOM driver (iXblue Photonics (exail): DR-VE-0.1-MO).

The initial round-trip length of the cavity, $l_{\text{cav}}$, was chosen to be $\SI{4.496}{\meter}$ in order to match the recoherence length of a 15~chips long sequence at $\SI{1}{\giga\hertz}$ PRN-modulation frequency. To fine adjust the matching between both length, the PRN-frequency was tuned by changing the reference clock frequency for the GTX-transceiver on the FPGA which is simply proportional. The cavity was resonant for a frequency of $f_{\text{PRN}}\,=\,\SI{0.998495}{\giga\hertz}$, meaning a length deviation of \SI{3.388}{\milli\meter} or \SI{0.15}{\percent} from the geometrically measured one.

Additionally, to other cases were set up and investigated. For one case the cavity length was matched to half the recoherence length, $l_{\text{cav}} / d_{\text{coh}} = 0.5$. This was achieved by tuning the PRN-frequency such that a 31-chip long sequence had a recoherence length of \SI{9.0047}{\meter} and thus double the optical cavity length. 

In the other case the cavity was matched to twice the recoherence length ($l_{\text{cav}} / d_{\text{coh}} = 2$) to demonstrate the possibility of using any integer multiple relation. For this case a shorter, 7-chip long sequence with a recoherence length around \SI{2.098}{\meter} was combined with a shortened cavity of twice the lengths. This shortening of the cavity was necessary due to limitations in the tuning range of the reference clock frequency driving the FPGA.

\subsubsection{Recording and treatment of data}
While for the first scans over the resonances of the cavities simply oscilloscope traces were recorded and compared, the step-wise measurement of relative transmitted power was done differently to gather more data in a simple way. For these measurements time series were recorded containing around 200 scans over the cavity's resonance with a saw-tooth ramp on the piezo-actuated mirror. From these time series the maximum transmitted powers reached in each individual scan were averaged and compared with the same measurement done without PRN-modulation active.

For the scanned measurement of the PRN-modulation resonance, the transmitted power and the current frequency of the FPGA-reference clock were recorded synchronously. From these recordings, the transmitted power in relation to PRN-detuning could be recovered.

The Pound-Drever-Hall error signals were again recorded as oscilloscope traces.


\providecommand{\noopsort}[1]{}\providecommand{\singleletter}[1]{#1}%

\end{document}